\documentclass[reprint,aps,prl,superscriptaddress]{revtex4-2}

\usepackage{bm,physics,amsmath,amssymb,
	graphicx,xcolor,embedfile,comment,mathrsfs,mathtools,xfrac}
\usepackage[utf8]{inputenc}
\usepackage[normalem]{ulem}
\usepackage{nicefrac, xfrac}

\usepackage[T1]{fontenc}
\usepackage{tikz}
\graphicspath{{figures/}}

\makeindex

\begin{document}

\title{Dispersive effects in ultrafast non-linear phenomena}

\author{Dusan Lorenc}
\affiliation{Institute of Science and Technology Austria,
    Am Campus 1, 3400 Klosterneuburg, Austria} 
    
\author{Zhanybek Alpichshev}
\email{alpishev@ist.ac.at}
\affiliation{Institute of Science and Technology Austria,
    Am Campus 1, 3400 Klosterneuburg, Austria} 
    
\begin{abstract}

{It is a basic principle that an effect cannot come before the cause. Dispersive relations that follow from this fundamental fact have proven to be an indispensable tool in physics and engineering. They are most powerful in the domain of linear response where they are known as Kramers-Kronig relations. However when it comes to nonlinear phenomena the implications of causality are much less explored, apart from several notable exceptions.
Here in this work we demonstrate how to apply the dispersive formalism to analyse the ultrafast nonlinear response in the context of the paradigmatic nonlinear Kerr effect. We find that the requirement of causality introduces a noticeable effect even under assumption that Kerr effect is mediated by quasi-instantaneous off-resonant electronic hyperpolarizability. We confirm this by experimentally measuring the time resolved Kerr dynamics in GaAs by means of a hybrid pump-probe Mach-Zehnder interferometer and demonstrate the presence of an intrinsic lagging between amplitude and phase responses as predicted by dispersive analysis. Our results describe a general property of the time-resolved nonlinear processes thereby highlighting the importance of accounting for dispersive effects in the nonlinear optical processes involving ultrashort pulses.}

\end{abstract}

\maketitle

\section{Introduction}
The amazing developments in ultrafast laser technology of the past few decades have now enabled probing materials at the unprecedented timescales down to those associated with quantum transition times between the stationary atomic states \cite{Zheltikov2021}. All the more important in these extreme regimes is to remain mindful of fundamental limitations imposed on the material response, both linear and nonlinear, by the basic physical principles. In the present work we explore the implications of causality on ultrafast nonlinear optical response in the context of the paradigmatic Kerr effect (KE), whereby the refractive index experienced by a probing beam depends on the intensity of light. This effect is one of the most important nonlinear optical effects, responsible for a great number of practically important phenomena such as self- and cross- phase modulation of the beams, Kerr lensing, self-focusing, optical soliton formation, optical switching and passive mode-locking \cite{Boyd2008, Diels2006}. Since most of these become particularly prominent in the context of ultrashort laser pulses, it can be said that Kerr effect owes most of its relevance to the fact that it can be often treated as an almost instantaneous interaction between the probing pulse and the pumping one, responsible for the refractive index changes. 

The mechanism of fast Kerr effect relies on off-resonant electronic transitions whose characteristic response times can be as short as several femtoseconds \citep{Boyd2008, Chang1981}. However, as we argue below, it would be a mistake to assume that quasi-instantaneous electronic response necessarily implies that the actual response of the probing beam to the pump intensity variations is also instantaneous. The reason for this is causality \cite{Bunge2017} which inevitably leads to the dispersion of response functions \cite{Toll1956}. This relation has been originally realized in the form of celebrated Kramers-Kronig relations \cite{Kronig1926, Kramers1927} and has since found applications in a broad array of subjects in physics \cite{Rouleau2013, Luo2021, Lovell1974, Harter2020, Roessler1965, Nussenzveig1972} and engineering \cite{Bode1945}. Unfortunately despite general character of the relationship between causality and  frequency dependence of response, the applications of dispersive analysis in nonlinear phenomena has been comparatively limited, save for several notable exceptions such as \cite{SheikBahae1990, Hutchings1992, Peiponen2004, Peiponen2009}.

In the present Letter we consider the Kerr effect mediated by quasi-instantaneous off-resonant electronic cloud polarization and show how to consistently apply Kramers-Kronig relations with time-dependent intensity $I(t)$. We apply this approach to the case of two gaussian pulses and demonstrate that dispersion introduces a considerable lag between the phase- and amplitude response of the probe. This lag is readily measurable with ultrafast pulses, therefore we proceed with an experimental realization in a time-resolved Mach-Zehnder interferometer in the model system GaAs. We confirm the presence of the dispersive lag with the magnitude and sign of the effect being consistent with theoretical expectations.

\section{Dispersive analysis of time-resolved Kerr effect} 
The essence of Kerr effect is the modification of linear refractive index of the given material in response to the square of applied electric field. In the case of optical-frequency fast alternating applied field, the correction $\delta n$ to the refractive index is proportional to the intensity $I$ of the optical beam: $\delta n  = n_2 I$, where the coefficient $n_2$ is referred to as either Kerr- or nonlinear refractive index. The realization that the end result of this {\it nonlinear} phenomenon (KE) is a correction to a {\it linear} response function (refractive index) has enabled a quantitative analysis of Kerr effect in terms of the classical Kramers-Kronig relations \cite{SheikBahae1990, Hutchings1992}. In particular, it was noted that when the intensity $I$ is constant, the refractive index acquires a constant correction which is usual Kramers-Kronig relations, automatically applies to the frequency dependence of the nonlinear refractive index $n_2(\omega)$.  

In many applications however, one often deals with time-dependent, pulse-shaped intensity profiles $I(t)$. In this case the analysis of the Kerr effect is not as straighforward anymore ({\it e.g.} one obviously cannot simply write $\delta n(t) = n_2 I(t)$). The goal of this section is to extend the dispersive analysis of Kerr effect of Sheik-Bahae {\it et al.} \cite{SheikBahae1990, Hutchings1992} to the case of ultrafast pulses. Although in many practical instances KE manifests itself the self interaction of a single beam, it is convenient to think about KE as an interaction between two distinct beams: pump and probe. Then KE can be viewed as the change in effective refractive index, experienced by probe beam $E_{pr}$, by the presence of the pump intensity $I_p$. For the purposes of this work we will ignore the spatial walk-off between the pump and probe beams due to group velocity dispersion. This can be justified for the case of thin sample or sufficiently close wavelength values of the two beams. Effectively this implies that we can ignore the spatial dispersion of light inside the sample. As demonstrated in the Supplementary Information, this is a valid assumption for the experimental settings used in this work. 

In the absence of spatial dispersion the general third-order nonlinear response of a system $\delta E_{NL}$ can be written in terms of the incident (probe) field $E(t)$ as:

\begin{equation}
\delta E_{NL}(t) = \int \displaylimits_{-\infty}^{t} dt_1 dt_2 dt_3 \,  \bar{\chi}^{(3)}(t; t_1, t_2, t_3)E(t_1) E(t_2) E(t_3)
\end{equation}

To remind, in this work we limit our scope to the case of Kerr effect: a special case of the general third-order nonlinearity where the phase and amplitude of probe field $E_{pr}(t)$ are modified in the presence of intensity of light $I_p(t)$ which, generally speaking, is time-dependent. In Kerr effect, the probe in a certain sense is oblivious to the phase of the pump beam. This sets KE apart from other third order nonlinear processes such as third harmonic generation or four-wave mixing. In this case we can simplify the equation above by introducing Kerr susceptibility $ \chi^K$ as follows:

\begin{equation}
\delta E_{NL}(t) = \int\displaylimits_{-\infty}^{t} dt' E_{pr}(t') \int\displaylimits_{-\infty}^{t} dt'' \chi^K (t, t', t'') I_p(t'')
\label{eq:chiK}
\end{equation} 

\begin{figure}
\includegraphics[scale=0.44]{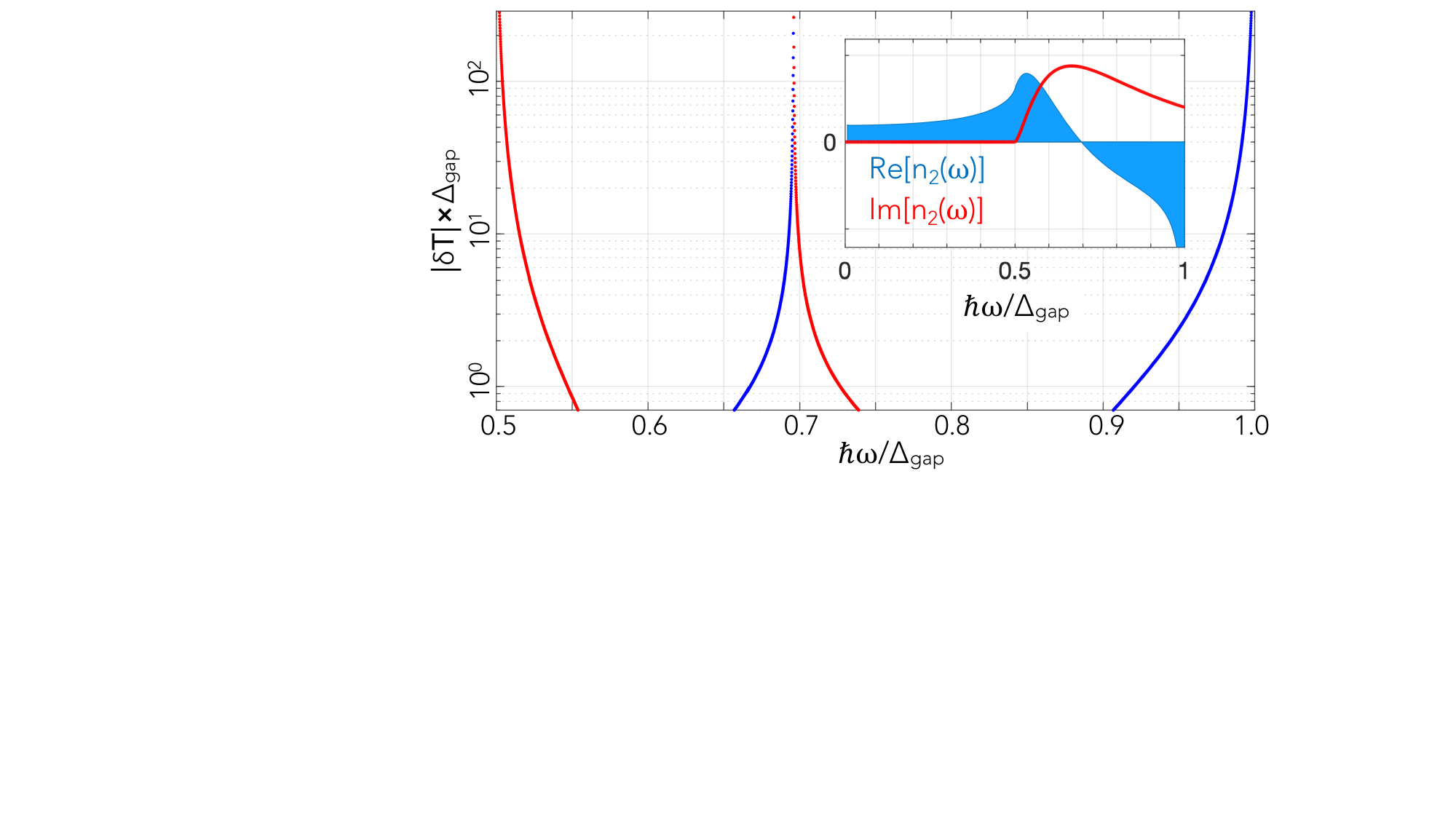}
\caption{Absolute value of phase-amplitude lag $|\delta T|$ in the units of $h/\Delta_g$ as a function of frequency as calculated from eq.(\ref{eq:lag}). Red (blue) segments correspond to positive (negative) lag with phase response lagging behind (ahead) of amplitude response. Inset: real (blue) and imaginary (red) parts of the degenerate nonlinear Kerr index $\bar{n}_2(\omega, \Omega=\omega)$ (see Refs.\cite{SheikBahae1990, Hutchings1992, SM})} 
\label{fig:SB}
\end{figure}

This dependence can be further simplified if Kerr response can be treated as instantaneous. This is approximately the case for the Kerr effect mediated by off-resonant electronic transitions, where response time of electrons is on the order of $\tau_{res}\sim h/\Delta_{gap}$, where $\Delta_{gap}$ is the band gap. The latter typically has a value on the order of 1eV, which makes $\tau_{res} \sim 4$fs, which is much shorter than the usual width of ultrafast pulses. If the main nonlinear interaction channel in the medium is the cooperative effect of pump and probe beams (such as 2PA or Raman effect), the nonlinear Kerr susceptibility $\chi^K$ can be written as:

\begin{equation}
\chi^K(t,t',t'')=f(t-t') \delta(t'-t'')
\label{eq:fdelta}
\end{equation}  

\noindent This is because in the absence of real (on-shell) intermediate excitations, the nonlinear interaction between pump and probe can only be present when both overlap in time. Now since $\chi^K(t, t', t'')=\chi^K(t-t', t-t'')=0$ whenever $t'$ or $t''$ precede $t$, $f(t)=0$ for $t<0$. Which means $f(t)$ is a causal function, and as such its Forurier transform must satisfy Kramers-Kronig relations. The equations above can be used to describe the Kerr effect with time-dependent pump intensity $I_p(t)$ as long as it doesn't vary too fast ($1/I_p\,(\partial I_p/\partial t)<\Delta_{gap}/h$). This condition is satisfied automatically whenever pump pulse can be described using slowly varying envelope approximation.

%
%


\noindent This way we reduce the full nonlinear susceptibility $\chi^{(3)}$ to a function of single variable $f(t)$. Importantly $f(t)=0$, whenever $t<0$, therefore it must be subject to Kramers-Kronig relations. In fact, the function $f(t)$ is directly related to the nonlinear Kerr index $\bar{n}_2(\omega, \Omega)$, where $\omega$ and $\Omega$ are the frequencies of the probe and pump beams respectively \cite{SM}: 

\def\niccefrac#1#2{    
 \raise.5ex\hbox{$#1$}%
 \kern-.1em/\kern-.15em%
  \lower.25ex\hbox{$#2$}}
\begin{equation}
f(\omega)=\int dt e^{i\omega t} f(t) \approx i \left({\omega L}/{c} \right)\bar{n}_2(\omega, \Omega)
\label{eq:relation}
\end{equation}

\noindent Combined, equations (\ref{eq:chiK}), (\ref{eq:fdelta}) and (\ref{eq:relation}) describe how to use dispersive relations with time-dependent pump intensity $I_p(t)$. As an example we can compute $\delta E_{NL}$ for the case of gaussian-shaped pump and probe pulses:
$$E_{pr}(t)=E_0 e^{-t^2/\tau^2}\cos(\Omega t),\,\,\mathrm{} \,\, I_p(t)= I_0 e^{-2(t-t_0)^2/\tau_1^2}$$
The modification of the (complex) refractive index of the medium implies that the phase and amplitude of the transmitted beam will acquire a change in phase $\Delta \phi$ and amplitude $\Delta A$ encoded in $\delta E_{NL}$. One can show that, as a function of delay $t_0$ between the pump and probe pulses, the amplitude and phase response will read (\cite{SM}):

\begin{align}  
\Delta A &\propto -I_0 \mathrm{Im}\{\bar{n}_2\} \exp \left[  -\frac{1}{\tilde{\tau}^2}  \left(   t_0 - \frac{  \nicefrac{\partial  \mathrm{Re}  \{\bar{n}_2\}  }{\partial \omega}}{\scriptstyle 2 \mathrm{Im} \{ \bar{n}_2 \}}    \right)^2 \right] \nonumber \\
\Delta \phi &\propto I_0 \mathrm{Re}\{\bar{n}_2\} \exp \left[  -\frac{1}{\tilde{\tau}^2}  \left(   t_0 + \frac{  \nicefrac{\partial  \mathrm{Im}  \{\bar{n}_2\}  }{\partial \omega}  }{\scriptstyle 2 \mathrm{Re} \{ \bar{n}_2 \}}    \right)^2 \right]
\label{eq:delays}
\end{align}

\noindent where $\tilde{\tau}^2=(\tau^2+\tau_1^2)/2$. According to these expressions, the dispersion of $\bar{n}_2$ produces a lag between the phase and amplitude response:

\begin{equation}
\delta T(\omega, \Omega)=\frac{1}{2} \left[ \frac{  \nicefrac{\partial  \mathrm{Re}  \{\bar{n}_2\}  }{\partial \omega}  }{ \scriptstyle  \mathrm{Im} \{ \bar{n}_2 \}}  +  \frac{  \nicefrac{\partial  \mathrm{Im}  \{\bar{n}_2\}  }{\partial \omega}  }{\scriptstyle  \mathrm{Re} \{ \bar{n}_2 \}} \right]
\label{eq:lag}
\end{equation}

The important feature of this lag is that it only depends on the dispersion of Kerr index $\bar{n}_2(\omega, \Omega)$ and is independent of pump intensity magnitude $I_0$. To find $\delta T(\omega, \Omega)$ we take the complex $\bar{n}_2(\omega, \Omega)$ as obtained by dispersive analysis assuming $\mathrm{Im}\{ \bar{n}_2 \}$ is mostly due to two-photon absorption (2PA)\cite{SM, Hutchings1992, SheikBahae1990}. In Fig.\ref{fig:SB} we plot ``degenerate'' $\delta T$  taken at $\omega=\Omega$. For reference we also plot the degenerate $\bar{n}_2(\omega, \Omega=\omega)$ in the inset of Fig.\ref{fig:SB}. As is evident from this figure, the lag is most prominent near one- and two-photon resonances. These frequencies correspond to special points $\bar{n}_2(\omega, \Omega)$ in the complex plane $\omega$ \cite{SM}. Somewhat less expected is the significant lag around $\hbar\omega\approx 0.7\Delta_{gap}$ which corresponds to the zero-crossing of $\mathrm{Re}\{ \bar{n}_2 \}$ (inset Fig.\ref{fig:SB}). Overall the analysis demonstrates that the lag $\delta T$ is quite significant especially near certain special points, therefore we are turning now to its experimental observation.

\begin{figure}
\includegraphics[scale=0.38]{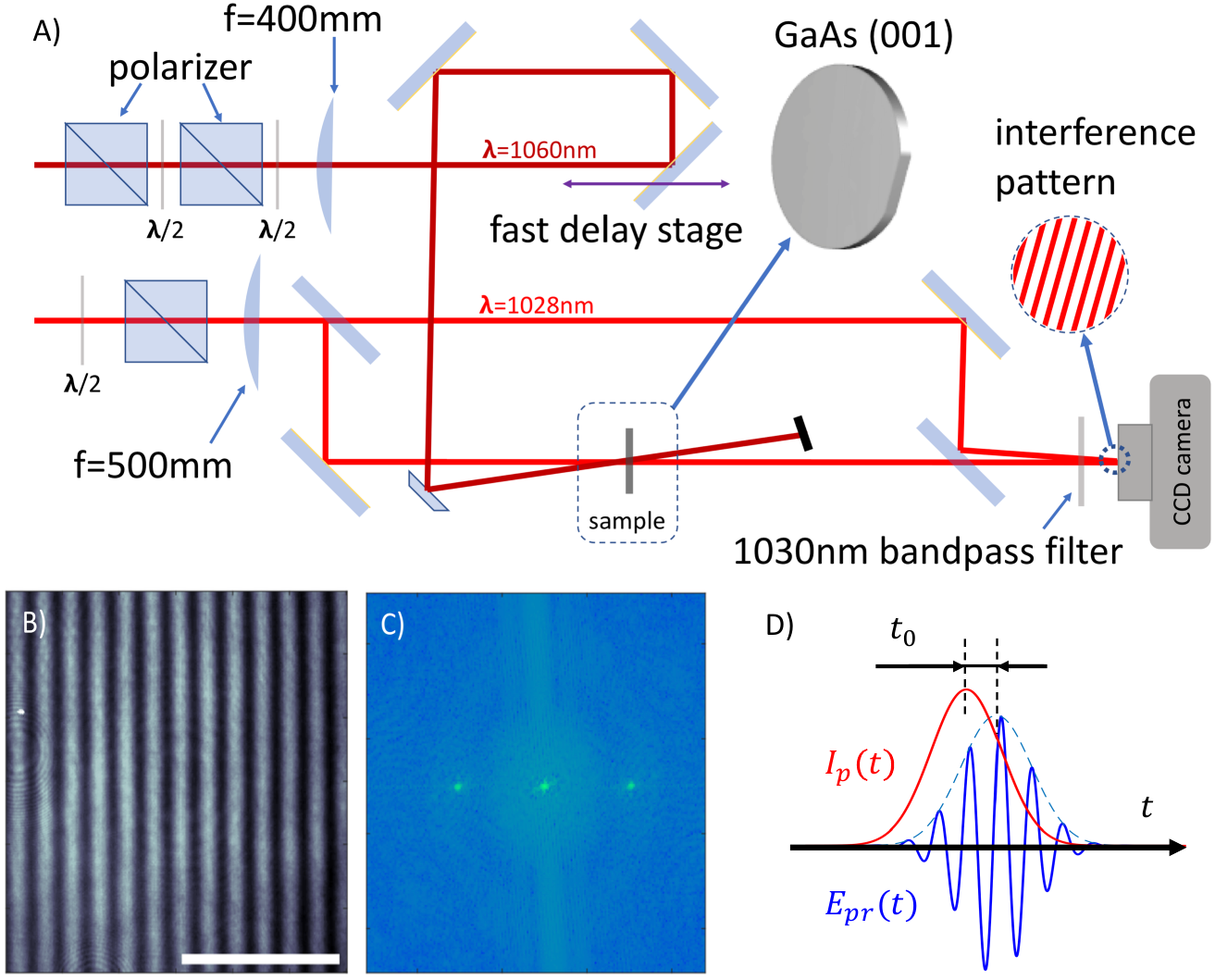}
\caption{A) Schematic layout of the time-resolved Mach-Zehnder interferometer. Pump and probe beams are near-degenerate ($\lambda=$ 1060nm and 1028nm respectively). Transient pump-induced changes in refractive index are detected as shift in the interference pattern recorded by a CCD camera; B) sample interferogram as recorded by CCD camera; C) 2D Fourier intensity of the interferogram in B). The real-space shift of the interference pattern is inferred from the change of the phase of complex amplitude at the peak. D) A schematic depiction of pump and probe pulses overlapping in time.}
\label{fig:interferometer}
\end{figure}

\section{Experiment} 

In order to experimentally observe the lagging effect, we study GaAs where Kerr effect is known to be dominated by electronic hyperpolarizability \cite{Chang1981}. To measure the nonlinear refractive index in a time-resolved manner we construct a pump-probe setup based on a Mach-Zehnder interferometer (MZI) as shown in Fig. \ref{fig:interferometer}A. Here one arm of the interferometer is the probe beam that passes through the GaAs sample,while the other is used a reference. Both beams are then combined on a CCD camera that records the interferogram. Pump is a separate beam with a slightly different wavelength and tunable delay $t_0$ with respect to probe. Changes in the real and imaginary parts of the refractive index induced by the pump, are recorded as respective changes in the position and amplitude of the interference pattern on the CCD matrix:

\begin{equation}
I({\bf r}, t) \propto |E_{pr}| |E_{ref}| \cos({\bf q} \cdot {\bf r} + \phi_g + 2 \pi n(t_0) L/\lambda)
\label{eq:interference}
\end{equation}

\noindent here $|E_{pr}|$ and $|E_{ref}|$ are the amplitudes of the electric field in the probe and reference arms; ${\bf q} = {\bf k_1} - {\bf k_2}$ is the difference between the corresponding beams' wave vectors ${\bf k_1}$ and ${\bf k_2}$;  $\phi_g$ is the static phase difference determined by interferometer geometry. The last term is the phase accumulated inside the sample with a thickness $L$ and refractive index $n$ which is in general a complex number. To extract the phase and amplitude information, we Fourier-transform each interferogram and evaluate the complex amplitude of the Fourier peak corresponding to the main peak in the interference pattern (Fig.\ref{fig:interferometer}B and C) $K(t_0)=|K(t_0)|e^{i \phi(t_0)}$ for each pump-probe delay $t_0$. The change in the absolute value and phase provide a direct measure of the pump-induced change in imaginary and real part of the refractive index: 

\begin{figure}[t]
\includegraphics[scale=0.4]{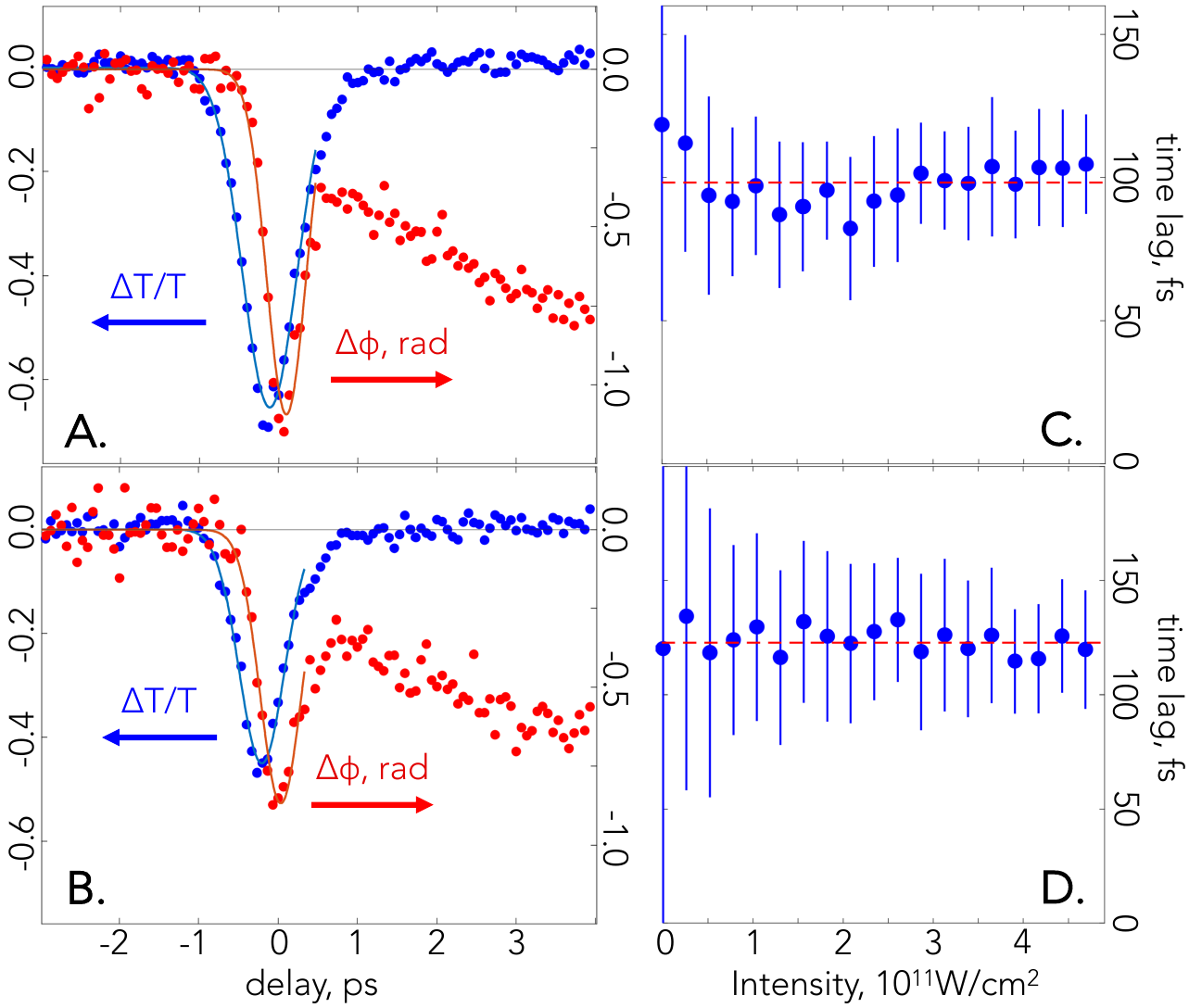}
\caption{A) Pump-induced amplitude (blue) and phase (red) change as a function of pump-probe delay for $xx$ configuration at $I_0=4.7$W$/$cm$^2$. Solid lines: gaussian fits to the peaks of corresponding colors; B) the same for $xy$ configuration; C) The phase-amplitude lag as a function of intensity $I_0$ is the $xx$- and; D)$xy$ configurations.}
\label{fig:GaAs_phase}
\end{figure}

\begin{figure}
\includegraphics[scale=0.37]{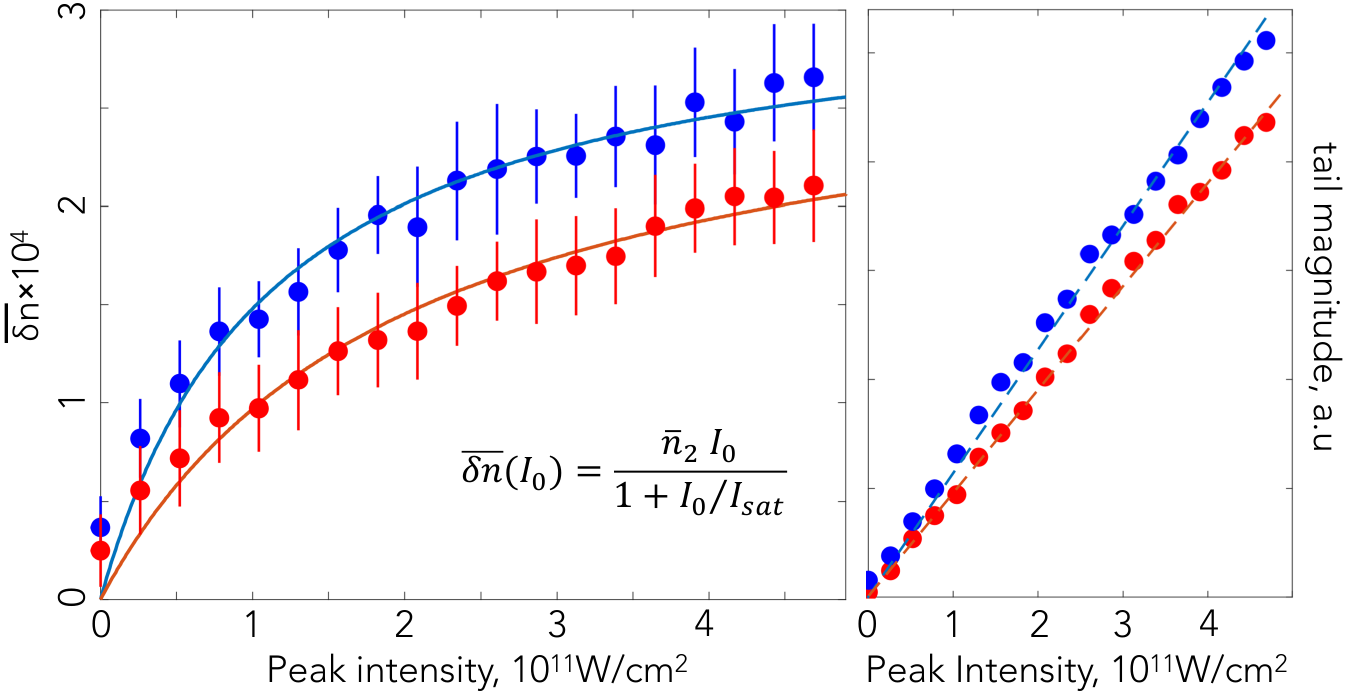}
\caption{A) Dependence of the effective instantaneous nonlinear refractive index $\overline{\delta n}$ on peak pump intensity $I_0$(see text). Solid lines are fits including saturation intensity $I_{sat}=0.8\times 10^{11}$W/cm$^2$, $\overline{\delta} n(I_0) = \overline{n}_2 I_0/(1+I_0/I_{sat})$; B) Pump intensity dependence of the long-lived tail in the phase response in Fig.\ref{fig:GaAs_phase}. In both panels the blue and red markers/lines correspond to $xx$ and $xy$ configurations respectively.}
\label{fig:GaAs_tail}
\end{figure}

The time dependence of pump-induced changes in phase and amplitude of the transmitted beam is shown in Fig.\ref{fig:GaAs_phase}. In Fig.\ref{fig:GaAs_phase}A the pump and probe are polarized parallel to the [100] crystalline axis of GaAs (``$xx$'' configuration) and in Fig.\ref{fig:GaAs_phase}B they are crossed (``$xy$''). The phase shift indicates that nonlinear Kerr index two distinct components: a fast peak and a slowly growing ``tail''. The quasi-instantaneous  character of the peak points towards off-resonant electronic polarizability as its origin, unlike the tail which will be discussed later. Using phase shift and sample thickness $L$ we can define an effective refractive index change as $\overline{\delta n} = \nicefrac{\Delta \phi}{2\pi} \cdot \nicefrac{\lambda}{L}$. Due to slight wavelength mismatch between pump and probe, this effective index cannot be directly used to accurately evaluate the Kerr index, however it can still provides a reasonable estimate (see \cite{SM}). With this in mind we plot $\overline{\delta n}$ against peak pump intensity in Fig.\ref{fig:GaAs_tail}A to find the values for the effective nonlinear Kerr index in both $xx$ and $xy$ configurations: $\overline{n}^{xx}_2=(-2.8 \pm 0.5) \times 10^{-15}$cm$^2$/W ; $\overline{n}^{xy}_2 = (-1.5\pm0.3)\times10^{-15}$cm$^2$/W.  Numeric simulations indicate that these numbers underestimate the actual Kerr indices by about a factor of 2 (\cite{SM}). Comparing these estimates against $n_2$ values measured at other wavelengths with similar pulse duration (e.g. $n_2(\lambda=1.7\mu$m$)\approx 3\times10^{-13}$cm$^2$/W \cite{Hurlbut2007}), shows that our working wavelengths $\lambda\approx 1\mu$m are close to the zero-crossing value of $\mathrm{Re}\{\bar{n}_2\}$ which implies that one should expect to see a significant lagging effect (see eq.(\ref{eq:lag})). 

Amplitude response $\Delta A$ is directly related to the transient differential absorption $\Delta T/T$ plotted in Fig.\ref{fig:GaAs_phase}A and Fig.\ref{fig:GaAs_phase}B in blue. Similarly to the main peak in phase response, $\Delta T/T$ exhibits quasi-instantaneous behavior and its sign is negative in agreement with 2PA. The fact that $n_2<0$ and $\lambda_{\mathrm{pump}}>\lambda_{\mathrm{probe}}$ (see Supplemental Material \cite{SM} and Fig.\ref{fig:interferometer}A) rules out coherent power transfer between probe and pump beams (``two-beam coupling'' \cite{Boyd2008}) as an alternative to 2PA as a mechanism for the probe amplitude variations.

Having established that temporal peaks in both $\Delta A$ and $\Delta \phi$ are coming from intrinsic electronic response of GaAs we finally turn to the delay between amplitude $\Delta T/T$ and phase $\phi$ responses that can be clearly seen in Fig.\ref{fig:GaAs_phase}A and Fig.\ref{fig:GaAs_phase}B. As shown in Fig.\ref{fig:GaAs_phase}C and Fig.\ref{fig:GaAs_phase}D the magnitude of the delays are $\delta T_{xx}\approx 100$fs and $\delta T_{xy}\approx 120$fs for the parallel and crossed configurations respectively. These values correspond to $\delta T\cdot \Delta_{gap}/h \approx 30$ which is consistent with the simplified model in Fig.\ref{fig:SB} (including the sign). Most importantly the lags do not show any dependence on pump intensity in line with the dispersive analysis. This observation rules out alternative lagging mechanisms based on nonlinear phenomena. This observation of a clear intensity-independent lag between phase and amplitude responses in agreement with theoretical expectations is the main result of this work. 

\section{Discussion} 
As mentioned above, in addition to the quasi-instantaneous peak, the phase response in Fig.\ref{fig:GaAs_phase} features a prominent long lived ``tail'' that is building up over the course several picoseconds and stays constant for the entire accessible delay range (see Fig.1 in Supplemental Material \cite{SM}). This tail clearly has a different origin as compared to the to the main peak. This can already be seen in the different dependence on pump intensity $I_p$ between the peak and tail shown in Fig.\ref{fig:GaAs_tail}. While the tail grows linearly with $I_p$, the peak exhibits a clear saturation with a characteristic intensity $I_{sat} = 8 \times 10^{10}$ W/cm$^2$ (both $xx$ and $yy$). Next, extended lifitime implies that the tail owes its origin to some real excitations produced by the pump. The linear pump intensity dependence (Fig. \ref{fig:GaAs_tail}B) excludes nonlinear absorption mechanisms, such as 2PA, as the possible origin of the tail; thermo-optic effect should also be excluded, since thermo-optic coefficient in GaAs is positive \cite{DellaCorte2000}. With these considerations we believe the most likely origin of the tail is the photo-excitation of the deep defect states (as previously reported in \cite{Biinas2011}) followed by long-lived low-density electron-hole plasma formation. The low plasma frequency ensures negative contribution to the refractive index. 

The presence of the tail illustrates the usefulness of time resolved methods for determining the intrinsic electronic Kerr index $n_2$. Our value for nonlinear Kerr index at $\lambda=1028$nm can be compared to a significantly larger one obtained in an earlier measurement at $1060$nm by Said et. al. \cite{Said1992}. The origin of the discrepancy likely lies in the fact that in \cite{Said1992} they were using 40ps laser pulses. At such timescales the tail will be giving an integrated contribution (see Fig.\ref{fig:GaAs_phase}A,B) that will overwhelm the intrinsic instantaneous response. Therefore we disregard this number in the present study and rely on the values obtained with ultrashort pulses as in \cite{Hurlbut2007}.

\begin{figure}
\includegraphics[scale=0.46]{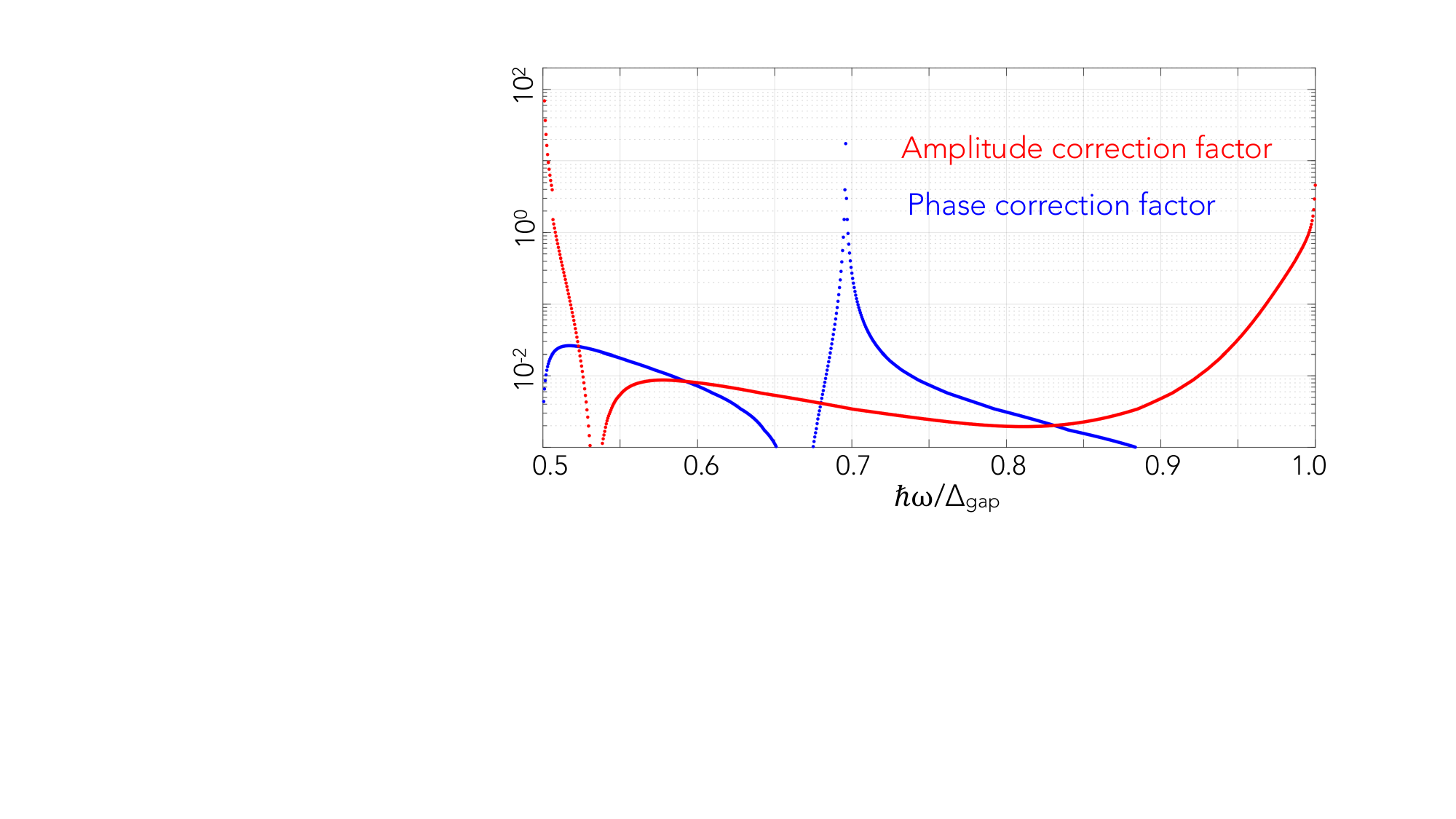}
\caption{Dispersion-induced renormalization factors for amplitude (blue) and phase (purple) response of a Kerr-active medium (see text).}
\label{fig:corrections}
\end{figure}

Finally, apart from the amplitude-phase lag the physics described above have further practically important implications. Indeed, the common case of a self-Kerr effect of a beam can be thought of as a special case of eq.(\ref{eq:delays}) with $t_0=0$. In this case as these equations indicate, the Kerr changes of both amplitude and phase will be systematically less than what one would expect without taking dispersive effects in account. Here we use the full expression for $\bar{n}_2(\omega, \Omega)$ (see \cite{SM}) to evaluate the correction factors $F_{\mathrm{ph}}$ and $F_{\mathrm{amp}}$ for phase and amplitude changes respectively, which we define as:

$$F_{ph} = |\Delta \phi_{\mathrm{max}}/\Delta \phi_0-1|,\,\,\, F_{amp} = |\Delta \mathrm{A}_{\mathrm{max}}/\Delta A_0-1|$$

\noindent where $\Delta \phi_{\mathrm{max}}$ and $\Delta \mathrm{A}_{\mathrm{max}}$ are the maximum phase and amplitude changes achievable by tuning $t_0$; while$\Delta \phi_0$ and $\Delta A_0$ are the same quantities evaluated at $t_0=0$. These correction factors show by how much the Kerr response of a system can exceed the nominal values calculated without regard to the dispersive effects discussed above. The correction factors calculated as function of frequency for a generic pulse duration $\tau=\tau_1=70$fs is plotted in Fig.\ref{fig:corrections}. As can be seen from the plot, the response renormalization becomes significant near the $\mathrm{Re}\{n_2\}$ sign-switching wavelength and near single- and two-photon resonances. 

\section{Conclusion} 
To summarize, we propose a procedure to extend the dispersive analysis of Sheik-Bahae et al. \cite{SheikBahae1990, Hutchings1992} to the situations where the pump intensity cannot be approximated as quasi-static and show that the resulting causal Kerr susceptibility $\chi_K(t, t', t'')$ can be directly related to the dispersive Kerr index $\bar{n}_2(\omega, \Omega)$. We find that dispersive effects can lead to significant modifications in phase and amplitude dynamics which lead to a significant underestimation of the peak Kerr response of a nonlinear medium. Our results shed light on the fundamental effects of causality for the ultrafast nonlinear phenomena and pave way for their efficient application.

\begin{acknowledgments}
\noindent \textbf{Acknowledgments.} The work was supported by the Institute of Science and Technology Austria (ISTA). We thank Prof. John M. Dudley, Dr. Ugur Sezer and Dr. Artem Volosniev for valuable discussions. 
\end{acknowledgments}

\bibliography{Bib_Kerr_lag}

\newpage\hbox{}\thispagestyle{empty}\newpage

\widetext

\section{Supplementary Information}

\noindent {\bf Experimental details.} The GaAs sample used in this study is a 675$\mu$m thick, (100)-oriented, semi-insulating GaAs wafer ( Freiberger Compound Materials GmbH). An amplified femtosecond laser system ({\it Light Conversion} Pharos) coupled with an optical parametric amplifier ({\it Light Conversion} Orpheus) is used as the principal laser source. The laser produces a train of pulses centred at 1028nm with a repetition rate of 3kHz, pulse duration of 290fs and a pulse energy of 2mJ. A small part of the beam (5\%) is sent to the Mach-Zehnder interferometer (MZI) through a variable attenuator consisting of a half-wave plate (HWP) and a Glan-Taylor polarizer; the remaining part of the beam pumps the optical parametric amplifier with a tunable output wavelength. 

In this work, the fundamental wavelength of $1028$nm is used as a probe while the 1060nm idler output from the optical parametric amplifier is used as the pump beam. The pump wavelength was chosen such that it is nearly degenerate with probe, but is still different enough to be effectively separable from the latter by means of a spectral filter. The delay between pump and probe is controlled with a motorized linear stage. Both pump and probe pass through a set of wave-plates and polarizers to attenuate their respective powers and to clean the polarizations. The pump beam is then passed through a additional HWP in order to control its polarization. The pump and probe beams are loosely focused and spatially overlapped in the sample in a slightly non-collinear arrangement. Care was taken to make sure that the probe diameter (D$_{\mathrm{probe}}\approx120\mu$m (1/$e^2$)) at the sample position is considerably smaller than that of the pump, the latter being evaluated as D$_{\mathrm{pump}}\approx 360\mu$m (1/$e^2$) by means of razor-blade method. This is made to ensure that pump intensity is a well defined quantity. The probe light leaving the output port of MZI passes through a $1030\pm5$nm bandpass filter and the resulting interferograms are recorded by a CCD camera ({\it PointGrey Research Inc.} CM3-U3-50S5M). The pump intensity was tuned in a broad range up to $\sim 5\times10^{11}$W/cm$^2$, while the probe intensity was kept fixed at $I_{\mathrm{probe}} = 4 \times 10^6$ W/cm$^2$. 

The pump-induced change in complex refractive index is obtained through measuring the changes in the phase $\phi(t_0)$ and amplitude $|K(t_0)|$ of the corresponding Fourier peak as function of delay $t_0$ between the pump and probe beams:

\begin{align*}
\mathfrak{Re} \{\Delta n(t_0)\}   &=  \frac{\lambda}{2\pi L} \left\{ \phi(t_0) - \phi(-\infty) \right\} \\
\mathfrak{Im}\{ \Delta n(t_0) \}  &\approx - \frac{\lambda}{2\pi L} \frac{ |K(t_0)| - |K(-\infty)| }{|K(-\infty)|}
\end{align*}

\noindent here by ``$t=-\infty$'' we mean a large enough negative delay when probe beam precedes the pump. The delay between the pump and probe pulses $t_0$ is controlled by a fast motorized delay stage (Newport LTLMS800). To extract the time dependence of transient changes to $n_2$ the stage is moved incrementally and an interferogram is recorded for each position of the stage. In order to avoid the detrimental effects of mechanical vibrations, for every delay stage position we record two consecutive interferograms: one at the actual desired position and the other at a fixed reference position pre-dating pump arrival (``$t=-\infty$''). The changes in both the amplitude and phase are found from the difference between the two interferograms. Given the high speed of the delay stage (500mm/s) this method helps minimize the effect of the low frequency noise coming from mechanical deformations of MZI. \\

\begin{figure}[h!]
\includegraphics[scale=0.5]{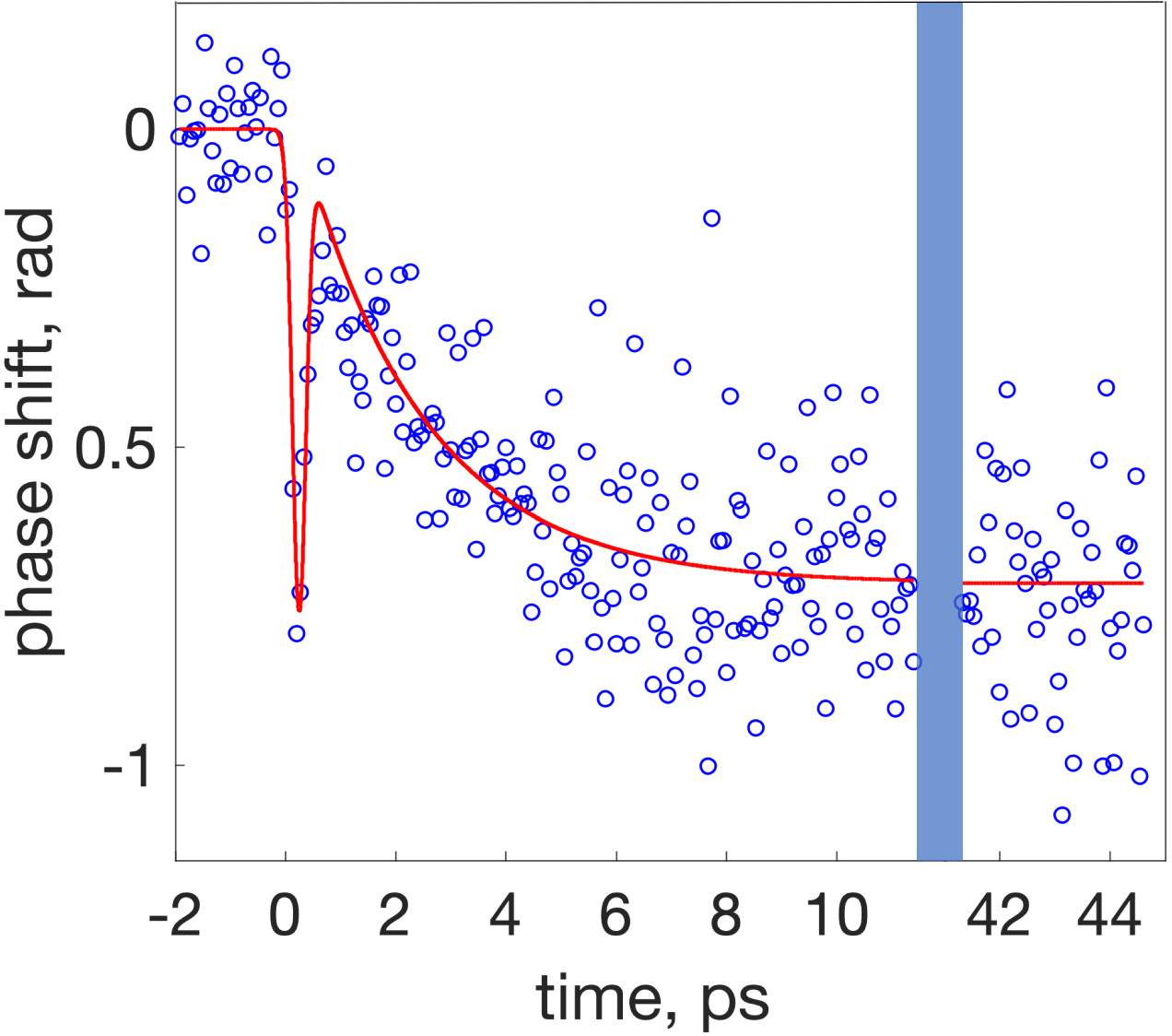}
\caption{Pump-induced phase shift in GaAs sample as a function of pump-probe delay in an extended range.} 
\label{fig:sm_long_tail}
\end{figure}

\noindent \textbf{Signal for large pump-probe delay values.} Here we study the temporal behavior of the ``tail'' whose onset is shown in Fig.3 of the main text. A long scan has been performed and the fig.\ref{fig:sm_long_tail}  shows that the signal remains virtually unchanged for pump-probe delays of multiple tens of picoseconds. On the one hand, the fact the feature has a long lifetime means that the pump is producing real excitation, on the other, the pump photon energy of 1.17eV which is less that GaAs band-gap energy of about $\Delta_{\mathrm{GaAs}}\approx$1.5eV. The most obvious mechanism for absorption of sub-gap photons is two-photon absorption. However this is ruled out by the scaling of the ``tail'' magnitude with pump intensity, which is linear, as can be seen in Fig.4B of the main text. Here it might be important to clarify, that this ruled-out real two-photon absorption is different from the virtual two photon process giving rise to the coherent feature around t=0 that is the main subject of this work. In the ruled-out two photon process both photons would be coming from the pump beam, while in the t=0 virtual process one of the two photons is still provided by the pump while the other comes from the probe. Therefore we can conclude the observed ``tail'' signal must originate from single-photon absorption. While such process is nominally prohibited for sub-gap frequencies, this is only strictly true for ideal samples. In realistic materials there always exist deep donor levels which can be excited with sub-gap photons. Similar processes have been reported reported previously in e.g. ref.\cite{Biinas2011}. \\

\noindent \textbf{Propagation of nearly degenerate pulses around $\lambda \approx 1\mu$m in GaAs.} The difference in wavelength between the pump and probe implies that the two have different group velocities. This gives rise to the walk-off between pulses as they propagate through the sample, which means that the effective interaction length is smaller than sample thickness $L$. In the main text the effective change in refractive index is defined as $\overline{\delta n} = (\Delta \phi / 2\pi)\cdot(\lambda / L)$ from which the effective Kerr index is inferred as $\overline{\delta n} = \overline{n}_2 I_0$, $I_0$ being (peak) pump pulse intensity. Now, since the interaction length is in fact less than $L$,  $\overline{n}_2 $ always underestimates the actual Kerr index value. Here we perform a simulation of pump-probe pulse interaction in order to find out the extent to which $\overline{n}_2 $ falls behind $n_2$ in magnitude when walk-off effects are taken into account.

For the present purpose we ignore absorption effects and go ahead with the standard equations for cross-phase modulation \cite{Agrawal}:

\begin{equation}
\frac{\partial A_j}{\partial z} +\beta_{1 j} \frac{\partial^2 A_j}{\partial t^2} = i n_2(\omega_j /c) \left\{  f_{jj} |A_j|^2 +2 f_{jk} |A_k|^2 \right\} A_j
\end{equation}

\noindent here the index $j=1,2$ corresponds to the probe and pump pulses respectively; $A_j(z,t)$ is the amplitude of $j$-th pulse, $z$ and $t$ being the distance in the propagation direction and time respectively; $\beta_{1j}$ and $\beta_{2j}$ are the group velocity and GVD coefficients of $j$-th pulse respectively; while $f_{jk}$ is the overlap coefficient (having the meaning of the inverse of the effective overlap area) between the $j$-th and $k$-th pulses with modal distributions $F_j(x,y)$:
$$
f_{jk} = \frac{\int dx dy  |F_j(x,y)|^2 |F_k(x,y)|^2}{\left( \int dx dy |F_j(x,y)|^2 \right)    \left(  \int dx dy |F_k(x,y)|^2 \right)}
$$

To compare with experiment the quantity of interest is $A_1(L,t)$, that is the amplitude of probe field at the exit from the sample. To mimic the experimental setup in which we measure the shifts of interference patterns, we first calculate the amplitude $A^0_1(L,t)$ after the sample in the absence of pump pulse. We will use it as a reference pulse playing the role of the pulse in the reference arm of the Mach-Zehnder interferometer if Fig.2 of the Main Text. Next, we add the pump pulse $A_2$. The temporal position of the pump in relation with probe (pump-probe delay) is kept as a free parameter $\tau_{\mathrm{delay}}$ and for each value, we calculate the field amplitude of the probe pulse that has interacted with the pump at the exit of the sample $A_1(L,t,\tau_{\mathrm{delay}})$.

In order to simulate the experimental interference pattern we calculate the ``interference'' between  sample field $A_1(L,t,\tau_{\mathrm{delay}})$ the reference field $A^0_1(L,t,\phi) \equiv A^0_1(L,t) \exp(i\phi)$ retarded by some variable phase $\phi$. The meaning of this phase is to simulate the actual interference pattern where e.g. depending on the point on the screen the probe and reference beams have different mutual phase dictated by the wavelength of the beams and their intersection angle. It is this space-dependent phase that gives rise to the interference pattern. With this in mind we calculate the cumulative intensity $G(\phi)$ as:

\begin{equation}
G(\phi, \tau_{\mathrm{delay}}) = \int dt \left( \left(A_1(L,t,\tau_{\mathrm{delay}})\right)^* A^0_1(L,t) \exp(i\phi) + {A_1(L,t,\tau_{\mathrm{delay}})} \left( A^0_1(L,t) \right)^* \exp(-i\phi) \right)
\end{equation}

\begin{figure}
\includegraphics[scale=0.6]{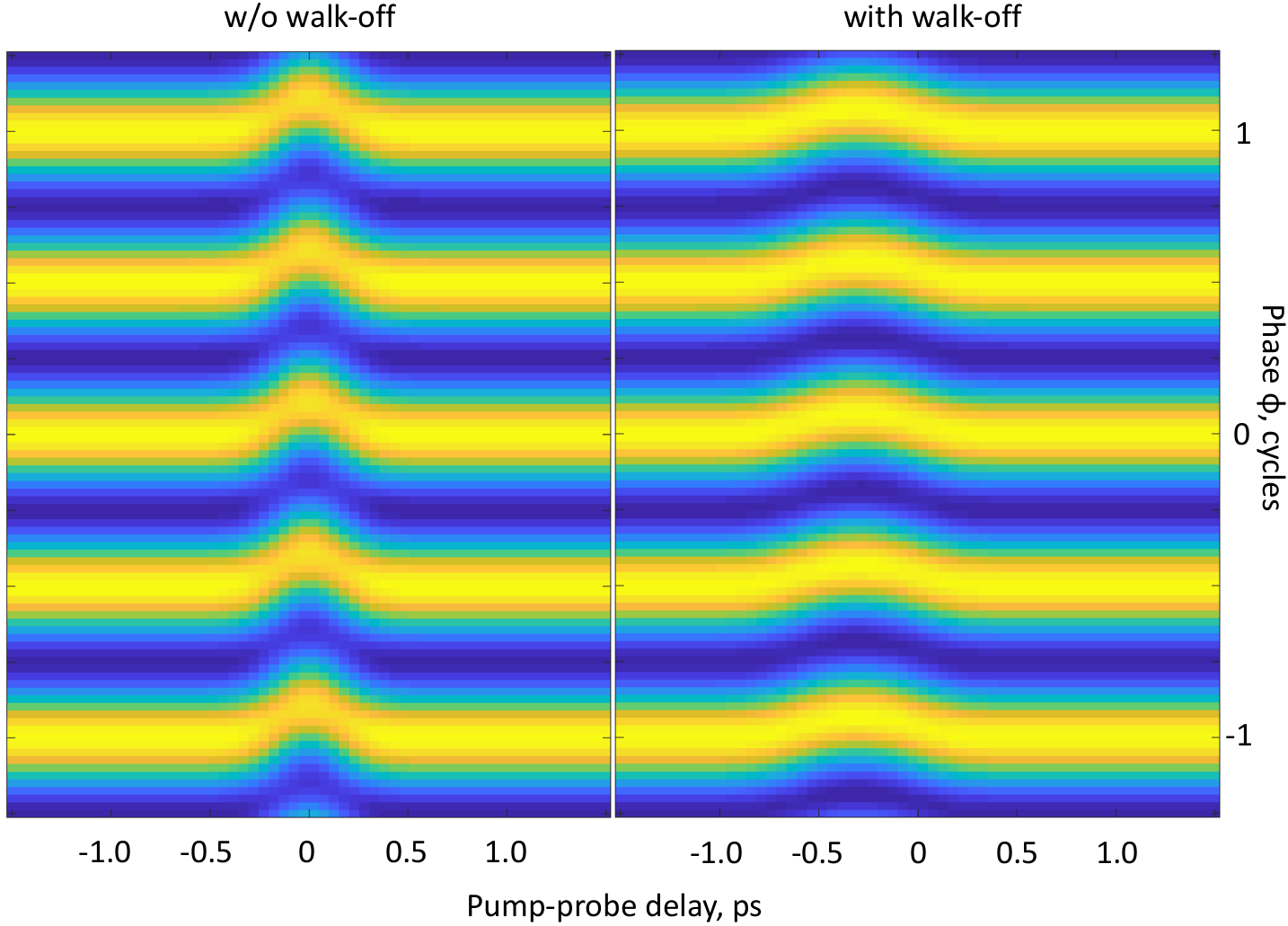}
\caption{``Interference patterns'' as a function of delay between pump and probe with (right) and without (left) group index mismatch between pump and probe pulses taken into account. Simulation was carried out at peak pump intensity $I_0 = 9.8\times 10^{14}$ W/m$^2$ and $n_2 = 1.2\times 10^{-19}$m$^2$/W. } 
\label{fig:walkoff_delay}
\end{figure}

In Fig.\ref{fig:walkoff_delay} we plot $G(\phi, \tau_{\mathrm{delay}})$ with and without group-index-mismatch effects taken into account. Here we use the following parameters:

\begin{itemize}
\item[]
Wavelength: $\lambda_{\mathrm{probe}}=1030$nm (probe);
$\lambda_{\mathrm{pump}}=1060$nm (pump);
\item[]
Pulse duration (pump and probe): $\tau_{\mathrm{pulse}}=270$fs
\item[]
Spot area (gaussian profile): A$_{\mathrm{eff1}}$ = $\pi(45\times10^{-6})^2$ m$^2$ (probe);
A$_{\mathrm{eff1}}$ = $\pi(180\times10^{-6})^2$ m$^2$ (pump);
\item[]
Sample thickness: $L=700\mu$m; Kerr nonlinear refractive index: $n_2=1.2\times 10^{-19}$ m$^2$/W
\item[]
Group refractive index: $n_{g1} = 3.9899$ (probe);
$n_{g2} = 3.9254$ (pump)(ref.~\cite{Skauli2003})
\item[]
GVD coefficient: $\beta_{11} = 4.5532$ (probe); 
 $\beta_{12} = 3.8172$ (pump) (ref.~\cite{Skauli2003});

\end{itemize}

\begin{figure}
\includegraphics[scale=0.6]{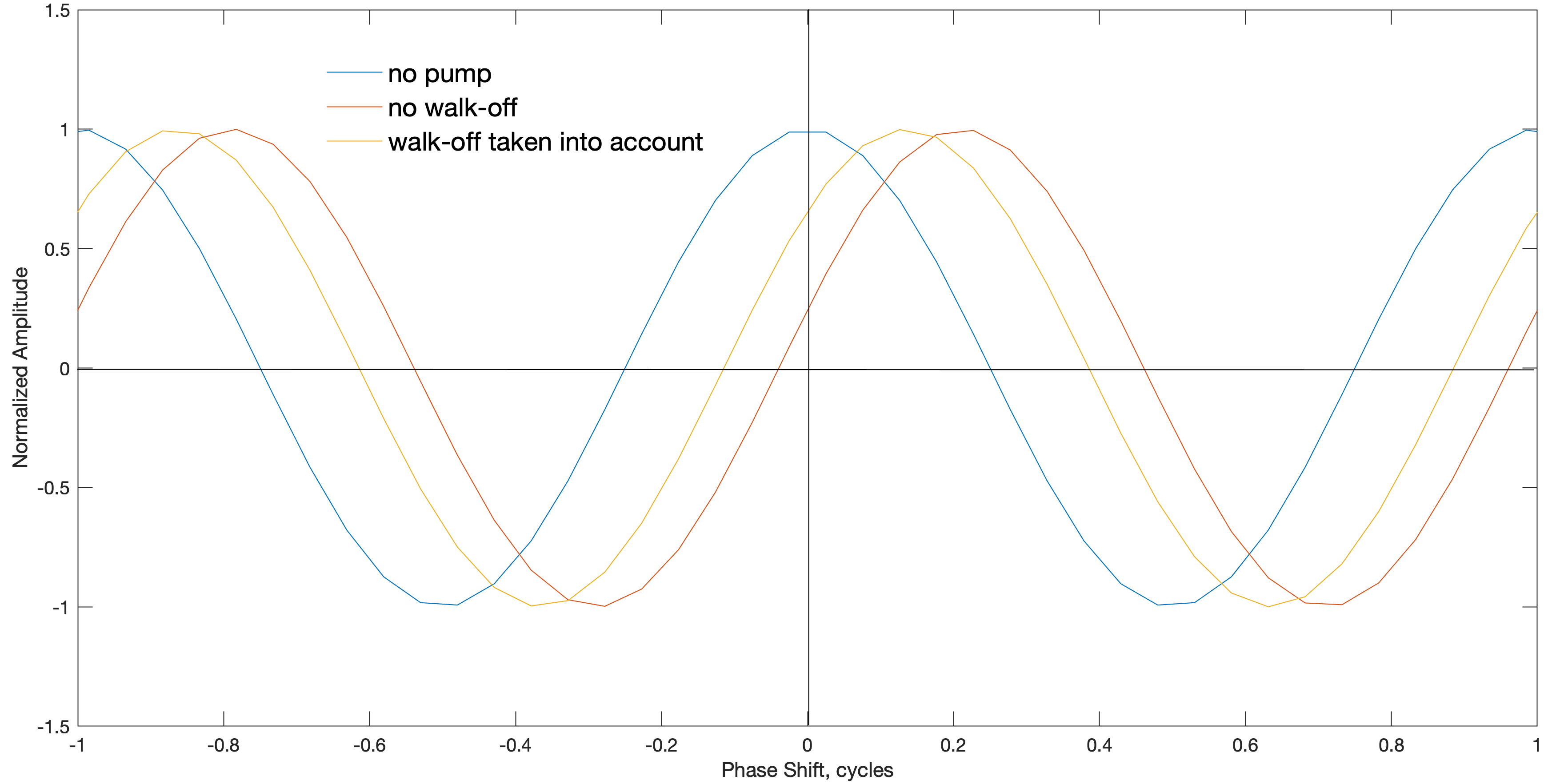}
\caption{Comparison of the ``interference patterns'' between the probe and reference beams (see main text of SM) in the absence of pump (blue); pump } 
\label{fig:walkoff}
\end{figure}

As can be seen, introducing the walk-off does not give give rise to an orders-of magnitude change in peak phase shift. To see this more clearly, in fig.\ref{fig:walkoff} we plot $G(\phi, \tau_{\mathrm{delay}})$ with $\tau_{\mathrm{delay}}$ fixed at maximum phase shift position. The peak phase shifts with and without walk-off compare as 0.65 to 1.
It is in place to mention that the peak phase $\Delta \phi \approx 1.2$rad shift in this simulation for $I_0 \approx 10^{11}$W/cm$^2$ is seemingly smaller than the experimentally measured $\Delta \phi_{\mathrm{exp}} \approx 0.5$rad, however this discrepancy is actually due to the fact that our simulation does not take saturation effects into account (see Fig.4 in the main text).

When walk-off effects are taken into account, the phase shift as can be seen in Fig.\ref{fig:walkoff} is reduced by about a factor of 2. This means that the ignoring the walk-off underestimates Kerr index only by the same factor of 2. This justifies the usage of the effective Kerr index in the main text for the purpose of the order-of-magnitude estimate of its actual value.

\subsection{Dispersive relations for quasi-instantaneous off-resonant electron-mediated Kerr effect}

\noindent \textbf{Relationship between Kerr susceptibility $\chi^K$ and nonlinear refractive index.} At this point it is important to establish the relation between the function $f(t)$ and nonlinear Kerr index $n_2$. With the assumptions made above one can express the field amplitude after the sample $E_{out}$ in terms of incoming field $E_{in}(t)=E_0\exp(-i \omega t)$ as follows:

\begin{equation}
E_{out}(t)=E_0\exp(-i \omega t)\times \exp(i\, 2\pi n_0(\omega) \nicefrac{L}{\lambda})
\end{equation} 

\noindent where $L$ is the sample thickness; $\lambda=(c\,\omega)/2\pi$ is the wavelength of the probing beam (in vacuum); and $n_0(\omega)$ is the linear refractive phase index of the medium which is in general a complex number. As was noted by Sheik-Bahae et al. (ref.\cite{SheikBahae1990}), when pump intensity $I_p$ is independent of time, the net result of nonlinear Kerr effect is a modification of the (linear) refractive index $n(\omega)$:

\begin{equation}
E_{out}(t)=E_0(t) \exp \bigl\{ i\, 2\pi \left[ n_0(\omega)+n_2(\omega, \Omega) I_p \right] \nicefrac{L}{\lambda} \bigr\}
\end{equation}

\noindent where $n_2(\Omega, \omega)$ is the nonlinear Kerr index of the sample material which in general depends on both pump ($\Omega$) and probe ($\omega$) frequencies. If we define $\delta E_{NL} = E_{out}(I_p) - E_{out}(I_p=0)$ as the nonlinear response of the medium, we can write:

\def\niccefrac#1#2{    
 \raise.5ex\hbox{$#1$}%
 \kern-.1em/\kern-.15em%
  \lower.25ex\hbox{$#2$}}
\begin{equation}
\delta E_{NL}(t) \approx E_0(t) \times \exp(i\, 2\pi n_0(\omega) \nicefrac{L}{\lambda}) \cdot \left( i \left(\niccefrac{\omega L}{c} \right)\bar{n}_2(\omega, \Omega) I_p \right)
\label{eq:smNL}
\end{equation}   

Noting that in practice it is usually the case that $n_0(\omega)$ varies by about $\sim 15$\% in the relevant frequency range of $\hbar \omega<\Delta_{gap}$ we follow Sheik-Bahae et al in refs.\cite{SheikBahae1990, Hutchings1992}, and ignore the frequency dependence of linear refractive index putting $n_0(\omega)=n_0$. Then the contribution of the linear term in the above expression is just a constant phase shift. In the experiment described in this work this term results in a constant phase offset of the beam in the probing arm of the interferometer independent of pump, which is subtracted away according to the measurement protocol. Alternatively one can imagine putting the same amount of sample material in the reference arm of the interferometer with no pump beam. Then the linear phase shift is equal in both arms and cancels out on the interferogram. Either way, the equilibrium linear phase shift is inconsequential for our analysis and will be dropped from now on. With this in mind we can re-write eq.(\ref{eq:smNL}) as:

\begin{equation}
\delta E_{NL}(\omega) =  i \left(\niccefrac{\omega L}{c} \right)\bar{n}_2(\omega, \Omega) I_p E_{pr}(\omega)
\label{eq:n2} 
\end{equation}

Now take eq.(\ref{eq:chiK}) and eq.(\ref{eq:fdelta}) and substitute $I_p(t)=I_0$. Integrating over delta-function we get:

\begin{equation*}
\delta E_{NL}(t)=\int \displaylimits_0^{t} dt' E_{pr}(t') f(t-t') I_0
\end{equation*}

\noindent and

\begin{equation}
\delta E_{NL}(\omega)=\int dt \exp(i \omega t) \delta E_{NL}(t)=f(\omega) E_{pr}(\omega) I_0 
\end{equation}

\noindent where $f(\omega) = \int dt \exp(i \omega t) f(t)$. Comparing this with eq.(\ref{eq:n2}) we see that

\begin{equation}
f(\omega)=i \left(\niccefrac{\omega L}{c} \right)\bar{n}_2(\omega, \Omega)
\label{eq:f2n}
\end{equation}

\noindent which is the relationship between $f(t)$ and nonlinear refractive (Kerr) index $\bar{n}_2(\Omega, \omega)$, where $\omega$ and $\Omega$ are the central frequencies of the probe and pump pulses respectively.\\

\noindent \textbf{Application to the case of two gaussian pulses.} To see implications of the formalism described in the previous paragraph, let's consider the case when both pump and probe have a gaussian profile shifted relative to each by $t_0$:
\begin{align}
I_p(t)=&\,I_0 \exp(-2(t-t_0)^2/\tau_1^2) \\
E_{pr}(t)=&\,E_0 \exp(-t^2/\tau^2) \cos(\omega_0 t)
\label{eq:Epr}
\end{align}

 \noindent then 
 
\begin{multline}
 \delta E_{NL}(t, t_0)=\int dt_1 f(t-t_1) I_p(t_1)E_{pr}(t_1)\propto\\
 \propto I_0 E_0 \int \frac{d\omega}{2\pi}f(\omega) \left\{    \exp \left[  \frac{\left(  2t_0/\tau_1^2+i(\omega+\omega_0)/2\right)^2 -\frac{2t_0^2}{\tau_1^2}-i\omega t  }{2/\tau_1^2+1/\tau^2}   \right]  + \exp \left[  \frac{\left(  2t_0/\tau_1^2+i(\omega-\omega_0)/2\right)^2 -\frac{2t_0^2}{\tau_1^2}-i\omega t  }{2/\tau_1^2+1/\tau^2}   \right] \right\}
 \label{eq:ENL}
\end{multline}
 
In practice, in order to characterize a rapidly oscillating pulse $E(t)$ one can interfere it with some reference pulse $E_\mathrm{ref}(t,\theta)$ with same central frequency as the characterized pulse, and a controllable phase $\theta$ relative to it. The measured signal is the total energy of the mixed pulse $\delta \mathcal E(\theta) \propto \int dt (E(t)+E_\mathrm{ref}(t,\theta))^2$. If the phase or amplitude of $E(t)$ changes for some reason, this will in general affect $\delta \mathcal E$.  More specifically, the phase and amplitude shifts of $E(t)$ are correspondingly proportional to:

\begin{align}
\Delta \phi \propto  \,\delta \mathcal E(\theta=0)& -  \delta \mathcal E(\theta=\pi)\nonumber \\
\Delta A \propto  \,\delta \mathcal E(\theta=\pi/2)& -  \delta \mathcal E(\theta=-\pi/2)
\label{eq:deltas}
\end{align}

Indeed, as an illustration one can imagine two laser beams being interfered at a screen while intersecting at some finite angle. Then if the amplitude of one of the beams has changed, this will affect the contrast ratio of the interference pattern, i.e. difference in intensity between the maximum (when mutual phase difference $\theta=0$) and minimum intensity positions (destructive interference, phase $\theta=\pi$). Likewise, the change in phase in one of the beams results in the shift of the interference pattern as a whole. The magnitude of the latter can be found as the difference in intensity changes at the positions of the highest spatial gradient of intensity on the pattern, i.e. when $\theta = \pm \pi/2$. 

The relevant part of $\delta \mathcal E(\theta)$ is the one that comes from the mixing of $E(t)$ and $E_\mathrm{ref}(t)$:

\begin{equation}
\delta \mathcal E_\mathrm{mix}(\theta)=2\int dt E(t) E_\mathrm{ref}(t)
\end{equation}

\noindent note that since $\theta$-dependence of $\delta \mathcal E(\theta)$ is entirely due to $\delta \mathcal E_\mathrm{mix}(\theta)$ one can substitute the former for the latter in eq.(\ref{eq:deltas}). 

In our experiment $E(t)=E_{pr}+\delta E_{NL}(t)$ with $E_{pr}$ and $\delta E_{NL}(t)$ defined in eq.(\ref{eq:Epr}) and eq.(\ref{eq:ENL}) respectively; $E_\mathrm{ref}(t, \theta)= E_{pr}(t+\theta/\omega_0)$. Now putting these into $\delta \mathcal E_\mathrm{mix}(\theta)$ we write:

\begin{align}
\begin{multlined}
\delta \mathcal E_\mathrm{mix}(t_0,\theta) = \mathrm{const}\cdot E_0^2 I_0 \int \displaylimits_{-\infty}^{+\infty} dt \int \displaylimits_{-\infty}^{+\infty} \frac{d\omega}{2\pi} f(\omega) \left[  e^{-\nicefrac{t^2}{\tau^2}+i\omega_0 t+ i\theta} +e^{-\nicefrac{t^2}{\tau^2}-i\omega_0 t- i\theta} \right] \cdot \\
 \cdot \left[  \exp\left\{  \frac{\left[ \nicefrac{2t_0}{\tau_1^2}+\nicefrac{i(\omega+\omega_0)}{2}  \right]^2}{\nicefrac{2}{\tau_1^2}+\nicefrac{1}{\tau^2}} - \frac{2t_0^2}{\tau_1^2} -i\omega t  \right\} 
 +\exp \left\{  \frac{\left[ \nicefrac{2t_0}{\tau_1^2}+\nicefrac{i(\omega-\omega_0)}{2}  \right]^2}{\nicefrac{2}{\tau_1^2}+\nicefrac{1}{\tau^2}} - \frac{2t_0^2}{\tau_1^2} -i\omega t  \right\}  \right]
\end{multlined}
\end{align} 

\noindent dropping the oscillating terms (rotating wave approximation) and integrating over $t$ gives:

\begin{multline}
\delta \mathcal E_\mathrm{mix}(t_0, \theta) \approx \mathrm{const}\cdot E_0^2 I_0 \int \frac{d \omega}{2\pi} f(\omega)\cdot\\
\cdot \left[ \exp\left\{ -\frac{2t_0^2}{\tau_1^2}-\left( \frac{\omega-\omega_0}{2}\right)^2\tau^2 +i\theta + 
\frac{\left[ 2 t_0/\tau_1^2+i(\omega - \omega_0)/2 \right]^2}{2/\tau_1^2+1/\tau^2}  \right\} +(\omega_0 \rightarrow -\omega_0, \theta \rightarrow -\theta) \right]
\end{multline}

This integral can be evaluated by using steepest descent method. However, noticing that as long as $\tau \gg h/\Delta_{gap}$, $f(\omega)$ is a slowly varying function as compared to the gaussian in the rectangular brackets above. Therefore the latter can be approximated as delta-function and the integral can then be evaluated trivially to give:

\begin{equation}
\delta \mathcal E_\mathrm{mix}(t_0, \theta) \approx \mathrm{const}\cdot E_0^2 I_0 \cdot e^{-\left(\nicefrac{t_0}{\bar{\tau}}\right)^2} \cdot \left[  f(\omega_0 + i \eta)e^{i\theta} +  f(-\omega_0 + i \eta)e^{-i\theta}  \right]
\end{equation} 

\noindent here we introduced $\bar{\tau}^2 = (\tau^2 + \tau_1^2)/2$ and $\eta =t_0/\bar{\tau}^2$. Now substituting this into eq.(\ref{eq:deltas}) and using eq.(\ref{eq:f2n})we can write down the amplitude $\Delta A$ and phase response $\Delta \phi$ as a function of pump-probe delay $t_0$ in terms of nonlinear refractive index $n_2(\omega)$ as:

\begin{align}
\Delta \phi (t_0) \propto  &\, I_0 \cdot \left[ - n_2(\omega_0+i\eta) + n_2(-\omega_0+ i\eta)   \right]  \nonumber \\
\Delta A(t_0) \propto  &\, I_0 \cdot i\left[ n_2(\omega_0 + i\eta) + n_2(-\omega_0 + i \eta)  \right]
\label{eq:deltas2}
\end{align}

For typical pulses $\eta\ll\omega_0$, therefore the expressions above can be evaluated to a good precision by means of Taylor expansion near $\pm \omega_0$. In general $n_2(\omega)$ is a complex-valued function of $\omega$ on a complex plane, however given that away from the branching points at $\omega_0=0.5\Delta_{gap}$ and $\omega_0=\Delta_{gap}$  $n_2(\omega)$ analytical and thus satisfies Cauchy-Riemann conditions. Then it is sufficient to know how $n_2(\omega)$ depends on $\omega$ on real axis alone to find $n_2$ values in its immediate vicinity:

\begin{align}
\Delta A(t_0) \propto &\, I_0 \, \mathrm{Im}\{ n_2(\omega_0)\}\cdot \exp \left[   -\frac{1}{\bar{\tau}^2} \left(  t_0 - \frac{\partial( \mathrm{Re} \{ n_2 \} )/ \partial \omega}{2 \mathrm{Im} \{ n_2 \} }  \right)^2  \right] \nonumber \\
\Delta \phi(t_0) \propto &\, I_0 \, \mathrm{Re}\{ n_2(\omega_0)\}\cdot \exp \left[   -\frac{1}{\bar{\tau}^2} \left(  t_0 + \frac{\partial( \mathrm{Im} \{ n_2 \} )/ \partial \omega}{2 \mathrm{Re} \{ n_2 \} }  \right)^2  \right]
\end{align}

\noindent the derivatives here are evaluated at $\omega = \omega_0$. In order to proceed further we need to know the functional dependence of the nonlinear refractive index $n_2(\omega)$ on frequency.\\

\noindent \textbf{Frequency dependence of nonlinear refractive index}. In this paragraph we write down for the record the expression for the nonlinear refractive index as calculated by means of the dispersive approach of Sheik-Bahae et al in refs.\cite{SheikBahae1990, Hutchings1992}. Generally speaking nonlinear refractive (Kerr) index is a function of both pump- and probe frequencies, $\Omega$ and $\omega$ respectively. In the cited papers the authors notice that when pump intensity is constant $I_p(t)=I_0$, the total Kerr-modified refractive index of the medium can still be treated as linear index when when used to describe the propagation of the probe beam. This means that, among other things, the correction to the refractive index $\delta n(\omega) = n_2(\omega, \Omega) I_0$ must satisfy Kramers-Kronig relations:

\begin{equation}
\mathrm{Re}\{ n_2(\omega, \Omega)\}=\frac{1}{\pi} \int \displaylimits_{\infty}^{+\infty} \frac{\mathrm{Im}\{ n_2(\omega', \Omega) \}}{\omega' - \omega} d\omega'
\label{eq:KKdef}
\end{equation}

Imaginary part of the Kerr index corresponds to pump-induced absorption of the probe beam. In the simplest case this can happen due to two-photon absorption (2PA). The magnitude for 2PA was calculated for a direct-gap semiconductor in refs.\cite{SheikBahae1991, Hutchings1992}. Being only interested in the frequency dependence of $n_2$ we write down the result of that work as:

\begin{equation}
\mathrm{Im}\{n_2(\omega, \Omega)\} = \mathrm{const} \cdot \frac{\left( \omega + \Omega - \Delta \right) ^{3/2}}{\omega^2\Omega^2 } \left( \frac{1}{\omega} + \frac{1}{\Omega} \right)^2 \cdot \Theta(\omega + \Omega - \Delta)
\end{equation}

\noindent where $\Theta(x)$ is the Heaviside step-function. Since in eq.(\ref{eq:KKdef}) we integrate along the entire real axis, we need to fix what $n_2$ is for negative $\omega$. The reality of $n_2(t-t')$ requires that $n_2(\omega) = n_2^*(-\omega)$. Therefore, the 2PA part of $n_2$ should read:

\begin{equation}
\begin{multlined}
\mathrm{Im}\{n^{\mathrm{2PA}}_2(\omega, \Omega)\}= \frac{\mathrm{const}}{\omega^4\Omega^4} \cdot \left\{ {\left( \omega + \Omega - \Delta \right) ^{3/2}}{} \left( {\omega} + {\Omega} \right)^2 \cdot \Theta(\omega + \Omega - \Delta) + \right. \\
\left. {+\left( \Omega - \omega - \Delta \right) ^{3/2}}{} \left({\Omega}- {\omega} \right)^2 \cdot \Theta( \Omega-\omega - \Delta)\right\}
\end{multlined}
\end{equation} 

Further, as will be clear below, we need to also symmetrize the above expression with respect to $\Omega$. These new terms achieved by substituting $\Omega \rightarrow -\Omega$ can be identified with Raman contribution to $n_2$:

\begin{gather}
\mathrm{Im}\{n^{\mathrm{2PA+Raman}}_2(\omega, \Omega)\} = \frac{\mathrm{const}}{\omega^4\Omega^4} \cdot \left\{ {\left( \omega + \Omega - \Delta \right) ^{3/2}}{} \left( {\omega} + {\Omega} \right)^2 \cdot \Theta(\omega + \Omega - \Delta) + \right. \nonumber \\
\left. {+\left( \Omega - \omega - \Delta \right) ^{3/2}}{} \left({\Omega}- {\omega} \right)^2 \cdot \Theta( \Omega-\omega - \Delta)+ \right.\nonumber \\
\left. +{\left( \omega - \Omega - \Delta \right) ^{3/2}}{} \left( {\omega} - {\Omega} \right)^2 \cdot \Theta(\omega - \Omega - \Delta) + \right. \nonumber \\
\left. {+\left( -\Omega - \omega - \Delta \right) ^{3/2}}{} \left( -{\Omega}- {\omega} \right)^2 \cdot \Theta( -\Omega-\omega - \Delta) \right\}
\end{gather} 

Now we can perform the integral in eq.(\ref{eq:KKdef}), however in addition to the physical branching points at $\omega=\pm\left( \Delta \pm \Omega \right)$ that correspond to certain two-photon absorption thresholds, there is an unphysical singularity at $\omega=0$. As discussed in ref.\cite{Hutchings1992} this is an artifact of the derivation of two-photon absorption probability by means of dipole (A$\cdot {\bf p}$) approximation. At low frequencies this singularity is cancelled by taking into account the omitted diamagnetic term A$^2$. Knowing that singularity at $\omega=0$ is fictitious, we can simply amend the expression above with a proper counter-term that will remove the divergence at $\omega=0$. 

\begin{figure}
\includegraphics[scale=0.38]{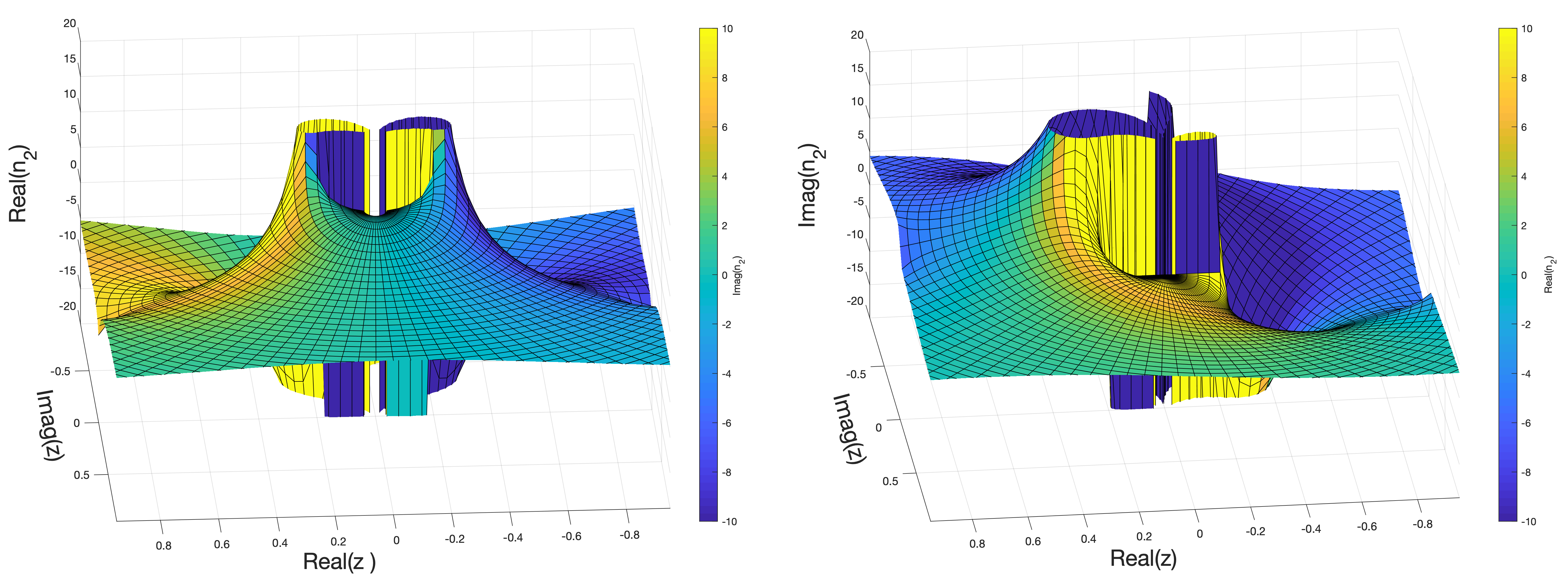}
\caption{Real(left) and imaginary (right) parts of $n_2(\omega, \Omega)$ as a function of complex $z=\omega/\Delta$ based of eq.(\ref{eq:full}). Here we took $\Omega = 0.8 \Delta$}
\label{fig:riemann}
\end{figure}

An inspection of the expression for non-degenerate two-photon absorption above reveals that it is symmetric in $\omega$ and $\Omega$. This is natural given how it is derived. What it means however, is that to recover proper $\omega$ dependence of $n_2$, one should also include the counter terms to remove the singularity at $\Omega=0$. Then one finally would need to make sure that counter terms in $\omega$ don't diverge at $\omega=0$ and vice versa. Adding everything up, and performing the Kramers-Kronig integration in eq.(\ref{eq:KKdef}) we can write down the full expression for nondegenerate $n_2$:

\begin{gather}
n^{\mathrm{2PA+Raman}}_2(\omega, \Omega) \propto \frac{\Delta^{9/2}}{\omega^4\Omega^4} \cdot \left\{ {\left(\Delta -\omega - \Omega   \right) ^{3/2}} \left( {\omega} +{\Omega} \right)^2  + \right. \nonumber 
\left. \left( \Delta + \omega - \Omega  \right) ^{3/2} \left( {\Omega}- {\omega} \right)^2 + \right.\nonumber \\
\left. +\left(\Delta  - \omega  + \Omega  \right) ^{3/2} \left( {\omega} -{\Omega} \right)^2  + \right. \nonumber 
\left. \left( \Delta + \Omega + \omega   \right) ^{3/2} \left( -{\Omega}- {\omega} \right)^2 - \right. \nonumber \\
\left. - 2 \Omega^2 (\Delta-\Omega)^{3/2} \left[   1+ \frac{\omega^2}{\Omega^2} - \frac{3\omega^2}{\Omega (\Delta -\Omega)} +\frac{3}{8} \frac{\omega^2}{(\Delta -\Omega)^2}  \right]- \right. \nonumber \\
\left. - 2 \Omega^2 (\Delta+\Omega)^{3/2} \left[   1+ \frac{\omega^2}{\Omega^2} + \frac{3\omega^2}{\Omega (\Delta +\Omega)} +\frac{3}{8} \frac{\omega^2}{(\Delta +\Omega)^2}  \right]- \right. \nonumber \\
\left. -2\omega^2 (\Delta - \omega)^{3/2} \left[  1+ \frac{\Omega^2}{\omega^2} - \frac{3 \Omega^2}{\omega (\Delta -\omega)} +\frac{3}{8} \frac{\Omega^2}{(\Delta - \omega)^2}  \right]- \right. \nonumber \\
\left. -2\omega^2 (\Delta + \omega)^{3/2} \left[  1+ \frac{\Omega^2}{\omega^2} + \frac{3 \Omega^2}{\omega (\Delta +\omega)} +\frac{3}{8} \frac{\Omega^2}{(\Delta + \omega)^2}  \right]+ \right. \nonumber \\
\left.+4\Delta^{3/2} \left( \omega^2 + \Omega^2 \right) + 9 \frac{\omega^2 \Omega^2}{\Delta^{1/2}} \right\}
\label{eq:full}
\end{gather} 

\noindent here both $\omega$ and $\Omega$ are understood as complex variables. In the degenerate limit $\omega=\Omega$ the above expression can be checked to coincide with the functional dependence of $n_2(\omega)$ given in \cite{Hutchings1992}.

\end{document}